\documentclass[12pt,a4paper]{article}%
\usepackage{amssymb}
\usepackage{amsfonts}
\usepackage{amsmath}
\usepackage{graphicx}
\usepackage[singlespacing]{setspace}
\usepackage{geometry}
\usepackage{url}%
\providecommand{\U}[1]{\protect\rule{.1in}{.1in}}

\begin{document}

\title{\textbf{A Simple Algorithm for Solving Ramsey Optimal Policy with Exogenous
Forcing Variables }}
\author{Jean-Bernard Chatelain\thanks{Paris School of Economics, Universit\'{e} Paris
I Pantheon Sorbonne, PjSE, 48 Boulevard Jourdan, 75014 Paris. Email:
jean-bernard.chatelain@univ-paris1.fr} and Kirsten Ralf\thanks{ESCE
International Business School, 10 rue Sextius Michel, 75015 Paris, Email:
Kirsten.Ralf@esce.fr.}}
\maketitle

\begin{abstract}
This algorithm extends Ljungqvist and Sargent (2012) algorithm of Stackelberg
dynamic game to the case of dynamic stochastic general equilibrium models
including exogenous forcing variables. It is based Anderson, Hansen,
McGrattan, Sargent (1996) discounted augmented linear quadratic regulator. It
adds an intermediate step in solving a Sylvester equation. Forward-looking
variables are also optimally anchored on forcing variables. This simple
algorithm calls for already programmed routines for Ricatti, Sylvester and
Inverse matrix in Matlab and Scilab. A\ final step using a change of basis
vector computes a vector auto regressive representation including Ramsey
optimal policy rule function of lagged observable variables, when the
exogenous forcing variables are not observable.

\textbf{JEL\ classification numbers}: C61, C62, C73, E47, E52, E61, E63.


\end{abstract}

\section{Introduction}

Ljungqvist and Sargent (2012, chapter 19) offer an elegant algorithm of
Stackelberg dynamic game used for Ramsey optimal policy. All dynamic
stochastic general equilibrium (DSGE) models include exogenous auto-regressive
forcing variables, which are not included in their algorithm. This algorithm
extends Ljungqvist and Sargent (2012, chapter 19) algorithm of dynamic
Stackelberg game to the case of DSGE\ models including exogenous forcing variables.

We use Anderson, Hansen, McGrattan, Sargent (1996) discounted augmented linear
quadratic regulator. After the usual algorithm for solving the Riccati
equation of the linear quadratic regulator (Amman (1996)), \emph{this
algorithm adds another step in solving a Sylvester equation for completing the
policy rule.} \emph{It also adds a term for the optimal initial anchor of
forward-looking variables on the predetermined forcing variables.}

This algorithm is easy to code and check. It is simple because it only calls
already optimized routines solving Ricatti and Sylvester equations and inverse
matrix in Matlab and Scilab. A\ final step using a change of basis vector
computes a vector auto regressive representation of Ramsey optimal policy. In
this representation of the Ramsey optimal policy rule, policy instruments
respond to lagged observable variables if all the exogenous forcing variables
are not observable.

\section{A\ Simple Algorithm}

\subsection{The Stackelberg problem}

We refer to Ljungqvist and Sargent (2012), chapter 19, step by step. The
Stackelberg leader is the government and the Stackelberg follower is the
private sector.

Let $\mathbf{k}_{t}$ be an $n_{k}\times1$ vector of controllable predetermined
state variables with initial conditions $\mathbf{k}_{0}$ given, $\mathbf{x}%
_{t}$ an $n_{x}\times1$ vector of endogenous variables free to jump at $t$
without a given initial condition for $\mathbf{x}_{0}$, and $\mathbf{u}_{t}$ a
vector of government policy instruments. Let $\mathbf{y}_{t}=(\mathbf{k}%
_{t}^{T},\mathbf{x}_{t}^{T})^{T}$ be an $\left(  n_{k}+n_{x}\right)  \times1$ vector.

Our only addition to Sargent and Ljungvist (2012) Stackelberg problem is to
include $\mathbf{z}_{t}$, which an $n_{z}\times1$ vector of non-controllable,
exogenous forcing state variables such as auto-regressive shocks. All
variables are expressed as absolute or proportional deviations about a steady state.

Subject to an initial condition for $\mathbf{k}_{0}$ and $\mathbf{z}_{0}$, but
not for $\mathbf{x}_{0}$, a government wants to maximize:%

\begin{equation}
-\frac{1}{2}%
{\displaystyle\sum\limits_{t=0}^{+\infty}}
\beta^{t}\left(  \mathbf{y}_{t}^{T}\mathbf{Q}_{yy}\mathbf{y}_{t}%
+2\mathbf{y}_{t}^{T}\mathbf{Q}_{yz}\mathbf{z}_{t}+\mathbf{u}_{t}%
^{T}\mathbf{Ru}_{t}\right)  \text{ }%
\end{equation}

where $\beta$ is the policy maker's discount factor and her policy preference
are the relative weights included matrices $\mathbf{Q,R}$\textbf{.
}$\mathbf{Q}_{yy}\geq\mathbf{0}$ is a $\left(  n_{k}+n_{x}\right)
\times\left(  n_{k}+n_{x}\right)  $ positive symmetric semi-definite matrix,
$\mathbf{R}>\mathbf{0}$ is a $p\times p$ \emph{strictly} positive symmetric
definite matrix so that policy maker's has at least a very small concern for
the volatility of policy instruments. The cross-product of controllable policy
targets with non-controllable forcing variables $\mathbf{y}_{t}^{T}%
\mathbf{Q}_{yz}\mathbf{z}_{t}$ is introduced by Anderson, Hansen, McGrattan
and Sargent (1996). To our knowledge, it has always been set to zero
$\mathbf{Q}_{yz}=\mathbf{0}$ so far in models of Ramsey optimal policy. This
simplifies the Sylvester equation in step 3.

The policy transmission mechanism of the private sector's behavior is
summarized by this system of equations written in a Kalman controllable
staircase form:%

\begin{equation}
\left(
\begin{array}
[c]{c}%
E_{t}\mathbf{y}_{t+1}\\
\mathbf{z}_{t+1}%
\end{array}
\right)  =\left(
\begin{array}
[c]{cc}%
\mathbf{A}_{yy} & \mathbf{A}_{yz}\\
\mathbf{0}_{zy} & \mathbf{A}_{zz}%
\end{array}
\right)  \left(
\begin{array}
[c]{c}%
\mathbf{y}_{t}\\
\mathbf{z}_{t}%
\end{array}
\right)  +\left(
\begin{array}
[c]{c}%
\mathbf{B}_{y}\\
\mathbf{0}_{z}%
\end{array}
\right)  \mathbf{u}_{t} \label{Transmission}%
\end{equation}

$\mathbf{A}$ is $\left(  n_{k}+n_{x}+n_{z}\right)  \times\left(  n_{k}%
+n_{x}+n_{z}\right)  $ matrix. $\mathbf{B}$ is the $\left(  n_{k}+n_{x}%
+n_{z}\right)  \times p$ matrix of the marginal effects of policy instruments
$\mathbf{u}_{t}$ on next period policy targets $\mathbf{y}_{t+1}$.

The government minimizes his discounted objective function by choosing
sequences $\left\{  u_{t},x_{t},k_{t+1},z_{t+1}\right\}  _{t=0}^{+\infty}$
subject to the policy transmission mechanism (\ref{Transmission}) and subject
to $2(n_{x}+n_{k}+n_{z})$ boundary conditions detailed below.

The certainty equivalence principle of the linear quadratic regulator (Simon
(1956)) allows us to work with a non stochastic model. "\textit{We would
attain the same decision rule if we were to replace }$x_{t+1}$\textit{ with
the forecast }$E_{t}x_{t+1}$\textit{ and to add a shock process }%
$C\varepsilon_{t+1}$\textit{ to the right hand side of the private sector
policy transmission mechanism, where }$\varepsilon_{t+1}$\textit{ is an i.i.d.
random vector with mean of zero and identity covariance matrix.}" (Ljungqvist
and Sargent, 2012 p.767).

The policy maker's choice can be solve with Lagrange multipliers using
Bellman's method (Ljungqvist and Sargent (2012)). It is practical (but not
necessary) to solve the policy maker's choice by attaching a sequence of
Lagrange multipliers $2\beta^{t+1}\mu_{t+1}$ to the sequence of private
sector's policy transmission mechanism constraints and then forming the Lagrangian:%

\begin{equation}
-\frac{1}{2}%
{\displaystyle\sum\limits_{t=0}^{+\infty}}
\beta^{t}\left[
\begin{array}
[c]{c}%
\mathbf{y}_{t}^{T}\mathbf{Q}_{yy}\mathbf{y}_{t}+2\mathbf{y}_{t}^{T}%
\mathbf{Q}_{yz}\mathbf{z}_{t}+\mathbf{u}_{t}^{T}\mathbf{Ru}_{t}+\\
2\beta^{t+1}\mathbf{\mu}_{t+1}\left(  \mathbf{A}_{yy}\mathbf{y}_{t}%
+\mathbf{B}_{y}u_{t}-\mathbf{y}_{t+1}\right)
\end{array}
\right]  \text{ }%
\end{equation}

The non-controllable variables dynamics can be excluded from the Lagrangian
(Anderson, Hansen, McGrattan and Sargent (1996)). It is important to partition
the Lagrange multipliers $\mathbf{\mu}_{t}$ conformable with our partition of
$\mathbf{y}_{t}=\left[
\begin{array}
[c]{c}%
\mathbf{k}_{t}\\
\mathbf{x}_{t}%
\end{array}
\right]  $, so that $\mathbf{\mu}_{t}=\left[
\begin{array}
[c]{c}%
\mathbf{\mu}_{k,t}\\
\mathbf{\mu}_{x,t}%
\end{array}
\right]  $, where $\mathbf{\mu}_{x,t}$ is an $n_{x}\times1$ vector of Lagrange
multipliers of forward-looking variables.

The first order conditions with the policy transmission mechanism leads to the
linear Hamiltonian system of the discrete time linear quadratic regulator
(Anderson, Hansen, McGrattan and Sargent (1996)).

$2(n_{x}+n_{k}+n_{z})$ boundary conditions determining the policy maker's
Lagrangian system with $2(n_{x}+n_{k}+n_{z})$ variables $(\mathbf{y}%
_{t},\mathbf{\mu}_{t},\mathbf{z}_{t})$ with $\mathbf{\mu}_{t}$ the policy
maker's Lagrange multipliers related to each of the controllable variables
$\mathbf{y}_{t}$ (table 1).

\textbf{Table 1:} $2(n_{x}+n_{k}+n_{z})$ boundary conditions%

\begin{tabular}
[c]{|c|c|}\hline
Number & Boundary conditions\\\hline
$n_{z}$ & $\underset{t\rightarrow+\infty}{\lim}\beta^{t}\mathbf{z}%
_{t}=\mathbf{z}^{\ast}=\mathbf{0}\text{, }\mathbf{z}_{t}\text{ bounded }%
$\\\hline
$+n_{k}+n_{x}$ & $\underset{t\rightarrow+\infty}{\lim}\beta^{t}\mathbf{y}%
_{t}=\mathbf{y}^{\ast}=\mathbf{0}\Leftrightarrow\underset{t\rightarrow+\infty
}{\lim}\frac{\partial L}{\partial\mathbf{y}_{t}}=\mathbf{0}=\underset
{t\rightarrow+\infty}{\lim}\beta^{t}\mathbf{\mu}_{t}\text{, }\mathbf{\mu}%
_{t}\text{ bounded}$\\\hline
$+n_{k}+n_{z}$ & $\mathbf{k}_{0}\text{ }$and $\mathbf{z}_{0}\text{
predetermined (given)}$\\\hline
$+n_{x}$ & $\mathbf{x}_{0}=\mathbf{x}_{0}^{\ast}\Leftrightarrow\text{ }%
\frac{\partial L}{\partial\mathbf{x}_{0}}=0=\mathbf{\mu}_{\mathbf{x}%
,t=0}^{\ast}\text{ predetermined}$\\\hline
\end{tabular}

Essential boundary conditions are the initial conditions of predetermined
variables $\mathbf{k}_{0}$ and $\mathbf{z}_{0}$ which are given.

Natural boundary conditions are such that the policy maker's anchors unique
optimal initial values of private sectors forward-looking variables. The
policy maker's Lagrange multipliers of private sector's forward (Lagrange
multipliers) variables are \emph{predetermined at the value zero:
}$\mathbf{\mu}_{\mathbf{x},t=0}=0$ in order to determine the unique optimal
initial value $\mathbf{x}_{0}=\mathbf{x}_{0}^{\ast}$ of private sector's
forward variables.

Bryson and Ho ((1975), p.55) explains natural boundary conditions as follows.
"\emph{If }$x_{t}$\emph{ is not prescribed at }$t=t_{0},$\emph{ it does not
follow that }$\delta x(t_{0})=0.$\emph{ In fact, there will be an optimum
value for }$x(t_{0})$\emph{ and it will be such that }$\delta L=0$\emph{ for
arbitrary small variations of }$x(t_{0})$\emph{ around this value. For this to
be the case, we choose }$\frac{\partial L}{\partial x(t_{0})}=\mu_{x,t_{0}}%
=0$\emph{ (1) which simply says that small changes of the optimal initial
value of the forward variables }$x(t_{0})$\emph{ on the loss function is zero.
We have simply traded one boundary condition: }$x(t_{0})$\emph{ given, for
another, (1). Boundary conditions such as (1) are sometimes called "natural
boundary conditions}" \emph{or transversality conditions associated with the
extremum problem.}"

Anderson, Hansen, McGrattan and Sargent (1996) assume a bounded discounted
quadratic loss function:%

\begin{equation}
E\left(
{\displaystyle\sum\limits_{t=0}^{+\infty}}
\beta^{t}\left(  \mathbf{y}_{t}^{T}\mathbf{y}_{t}+\mathbf{z}_{t}^{T}%
\mathbf{z}_{t}+\mathbf{u}_{t}^{T}\mathbf{u}_{t}\right)  \right)
<+\infty\text{ }%
\end{equation}

This implies a stability criterion for eigenvalues of the dynamic system such
that $\left\vert \left(  \beta\lambda_{i}^{2}\right)  ^{t}\right\vert
<\left\vert \beta\lambda_{i}^{2}\right\vert <1$, so that stable eigenvalues
are such that $\left\vert \lambda_{i}\right\vert <1/\sqrt{\beta}<1/\beta$. A
preliminary step is to multiply matrices by $\sqrt{\beta}$as follows
$\sqrt{\beta}\mathbf{A}_{yy}$\ $\sqrt{\beta}\mathbf{B}_{y}$ in order to apply
formulas of Riccati and Sylvester equations for the non-discounted augmented
linear quadratic regulator (Anderson, Hansen, McGrattan and Sargent (1996)).

\subsection{Preliminary step: Check if the system is stabilizable}

\textbf{Assumption 1:} The matrix pair ($\sqrt{\beta}\mathbf{A}_{yy}$%
\ $\sqrt{\beta}\mathbf{B}_{y}$) is controllable (all forward-looking variables
are controllable).

The matrix pair ($\sqrt{\beta}\mathbf{A}_{yy}$\ $\sqrt{\beta}\mathbf{B}_{y}$)
is controllable if the Kalman (1960) controllability matrix has full rank:
\begin{equation}
\text{rank }\left(  \sqrt{\beta}\mathbf{B}_{y}\text{ \ }\beta\mathbf{A}%
_{yy}\mathbf{B}_{y}\text{ \ }\beta^{\frac{3}{2}}\mathbf{A}_{yy}^{2}%
\mathbf{B}_{y}\text{ \ ... \ }\beta^{\frac{n_{k}+n_{x}}{2}}\mathbf{A}%
_{yy}^{n_{k}+n_{x}-1}\mathbf{B}_{y}\right)  =n_{k}+n_{x}%
\end{equation}

\textbf{Assumption 2:} The system is stabilizable when the transition matrix
$\mathbf{A}_{zz}$ for the non-controllable variables has stable eigenvalues,
such that $\left\vert \lambda_{i}\right\vert <1/\sqrt{\beta}$.

\subsection{Step 1: Stabilizing solution of a linear quadratic regulator}

"\emph{Step 1 and 2\ seems to disregard the forward-looking aspect of the
problem (step 3 will take account of that). If we temporarily ignore the fact
that the }$x_{0}$\emph{ component of the state }$y_{0}$\emph{ is not actually
a state vector, then superficially the Stackelberg problem has the form of an
optimal linear regulator.}" (Ljungqvist and Sargent (2012, p.769)).

When the forcing variables are set to zero $\mathbf{z}_{t}=\mathbf{0}$, a
stabilizing solution of the linear quadratic regulator satisfies:%
\begin{equation}
\mathbf{\mu}_{t}=\mathbf{P}_{y}\mathbf{y}_{t}%
\end{equation}
where $\mathbf{P}_{y}$ solves the matrix Riccati equation (Anderson, Hansen,
McGrattan and Sargent (1996)):%

\begin{equation}
\mathbf{P}_{y}\mathbf{=}\mathbf{Q}_{y}+\beta\mathbf{A}_{yy}^{^{\prime}%
}\mathbf{P}_{y}\mathbf{A}_{yy}-\beta^{^{\prime}}\mathbf{A}_{yy}^{^{\prime}%
}\mathbf{P}_{y}\mathbf{B}_{y}\left(  \mathbf{R}+\beta\mathbf{B}_{y}^{^{\prime
}}\mathbf{P}_{y}\mathbf{B}_{y}\right)  ^{-1}\beta\mathbf{B}_{y}^{^{\prime}%
}\mathbf{P}_{y}\mathbf{A}_{yy}%
\end{equation}

The optimal rule of the linear quadratic regulator is:%
\begin{equation}
\mathbf{u}_{t}=\mathbf{F}_{y}\mathbf{y}_{t}%
\end{equation}
where $\mathbf{F}_{y}$ is computed knowing $\mathbf{P}_{y}$ (Anderson, Hansen,
McGrattan and Sargent (1996)):%

\begin{equation}
\mathbf{F}_{y}=\left(  \mathbf{R+}\beta\mathbf{B}_{y}^{^{\prime}}%
\mathbf{P}_{y}\mathbf{B}_{y}\right)  ^{-1}\beta\mathbf{B}_{y}^{\prime
}\mathbf{P}_{y}\mathbf{A}_{yy}%
\end{equation}

As demonstrated by Simon (1956) certainty equivalence principle and by Kalman
(1960) solution, the optimal rule parameters $\mathbf{F}_{y}$ and
$\mathbf{P}_{y}$ of the linear quadratic regulator are independent of additive
random shocks and of initial conditions. This confirms that it is correct to
temporarily ignore the fact that $\mathbf{x}_{0}$ is not a state vector.

\subsection{Step 2: Stabilizing solution of an augmented linear quadratic
regulator}

This is the additional step missing in Ljungqvist and Sargent (2012)
algorithm. A stabilizing solution of the augmented linear quadratic regulator
satisfies (Anderson, Hansen, McGrattan and Sargent (1996)):%
\begin{equation}
\mathbf{\mu}_{t}=\mathbf{P}_{y}\mathbf{y}_{t}+\mathbf{P}_{z}\mathbf{z}_{t}%
\end{equation}
where $\mathbf{P}_{z}$ solves the matrix Sylvester equation:%

\begin{equation}
\mathbf{P}_{z}=\mathbf{Q}_{yz}+\beta\left(  \mathbf{A}_{yy}+\mathbf{B}%
_{y}\mathbf{F}_{y}\right)  ^{\prime}\mathbf{P}_{y}\mathbf{A}_{yz}+\beta\left(
\mathbf{A}_{yy}+\mathbf{B}_{y}\mathbf{F}_{y}\right)  ^{\prime}\mathbf{P}%
_{z}\mathbf{A}_{zz}%
\end{equation}

The optimal rule of the augmented linear quadratic regulator is:%
\begin{equation}
\mathbf{u}_{t}=\mathbf{F}_{y}\mathbf{y}_{t}+\mathbf{F}_{z}\mathbf{z}_{t}%
\end{equation}
where $\mathbf{F}_{z}$ is computed knowing $\mathbf{P}_{z}$:%

\begin{equation}
\mathbf{F}_{z}=\left(  \mathbf{R+}\beta\mathbf{B}_{y}^{^{\prime}}%
\mathbf{P}_{y}\mathbf{B}_{y}\right)  ^{-1}\beta\mathbf{B}_{y}^{\prime}\left(
\mathbf{P}_{y}\mathbf{A}_{yz}+\mathbf{P}_{z}\mathbf{A}_{zz}\right)
\end{equation}

As demonstrated by Simon (1956) certainty equivalence principle and by
Anderson, Hansen, McGrattan and Sargent (1996) solution, the optimal rule
parameters $\mathbf{F}_{z}$ and $\mathbf{P}_{z}$ of the augmented linear
quadratic regulator are independent of additive random shocks and of initial
conditions. This confirms that it is correct to temporarily ignore the fact
that $\mathbf{x}_{0}$ is not a state vector, until step 3.

\subsection{Step 3: Solve for $x_{0}$, the optimal initial anchor of
forward-looking variables}

The policy maker's Lagrange multipliers on private sector forward-looking
variables are such that $\mathbf{\mu}_{0,x}=\mathbf{0}$, at the initial date.
The optimal stabilizing condition is:%
\begin{equation}
\left(
\begin{array}
[c]{c}%
\mathbf{\mu}_{0,k}\\
\mathbf{\mu}_{0,x}%
\end{array}
\right)  =\left(
\begin{array}
[c]{cc}%
\mathbf{P}_{y,k} & \mathbf{P}_{y,kx}\\
\mathbf{P}_{y,kx} & \mathbf{P}_{y,x}%
\end{array}
\right)  \left(
\begin{array}
[c]{c}%
\mathbf{k}_{0}\\
\mathbf{x}_{0}%
\end{array}
\right)  +\left(
\begin{array}
[c]{c}%
\mathbf{P}_{z,k}\\
\mathbf{P}_{z,x}%
\end{array}
\right)  \mathbf{z}_{0}=\left(
\begin{array}
[c]{c}%
\mathbf{\mu}_{0,k}\\
\mathbf{0}%
\end{array}
\right)
\end{equation}

This implies%

\begin{equation}
\mathbf{P}_{y,kx}\mathbf{k}_{0}+\mathbf{P}_{y,x}\mathbf{x}_{0}+\mathbf{P}%
_{z,x}\mathbf{z}_{0}=\mathbf{0}%
\end{equation}

Which provides the optimal initial anchor:%

\begin{equation}
\mathbf{x}_{0}=\mathbf{P}_{y,x}^{-1}\mathbf{P}_{y,kx}\mathbf{k}_{0}%
+\mathbf{P}_{y,x}^{-1}\mathbf{P}_{z,x}\mathbf{z}_{0}%
\end{equation}

The exogenous forcing variables adds the term $\mathbf{P}_{y,x}^{-1}%
\mathbf{P}_{z,x}\mathbf{z}_{0}$ with respect to Ljungqvist and Sargent (2012) algorithm.

\subsection{Step 4: Compute impulse response functions and optimal loss
function}

The transmission mechanism is given. Computing $\mathbf{F}_{y}$ and
$\mathbf{F}_{z}$ provides a reduced form of the optimal policy rule. Computing
$\mathbf{P}_{y}$ and $\mathbf{P}_{z}$ provides the missing initial conditions.%

\begin{align*}
\left(
\begin{array}
[c]{c}%
E_{t}\mathbf{y}_{t+1}\\
\mathbf{z}_{t+1}%
\end{array}
\right)   &  =\left(
\begin{array}
[c]{cc}%
\mathbf{A}_{yy} & \mathbf{A}_{yz}\\
\mathbf{0}_{zy} & \mathbf{A}_{zz}%
\end{array}
\right)  \left(
\begin{array}
[c]{c}%
\mathbf{y}_{t}\\
\mathbf{z}_{t}%
\end{array}
\right)  +\left(
\begin{array}
[c]{c}%
\mathbf{B}_{y}\\
\mathbf{0}_{z}%
\end{array}
\right)  \mathbf{u}_{t}\\
\mathbf{u}_{t}  &  =\mathbf{F}_{y}\mathbf{y}_{t}+\mathbf{F}_{z}\mathbf{z}%
_{t}\\
\mathbf{x}_{0}  &  =\mathbf{P}_{y,x}^{-1}\mathbf{P}_{y,kx}\mathbf{k}%
_{0}+\mathbf{P}_{y,x}^{-1}\mathbf{P}_{z,x}\mathbf{z}_{0}\text{, }%
\mathbf{k}_{0}\text{ and }\mathbf{z}_{0}\text{ given}%
\end{align*}

This information is sufficient to compute impulse response functions (the
optimal path of the expected values of variables $\mathbf{y}_{t}$
$\mathbf{z}_{t}$ and $\mathbf{u}_{t}$) and to sum up over time their value in
the the discounted loss function.

By contrast to other algorithms based on Miller and Salmon (1985) solution,
\emph{it is not necessary} to compute all the values over time of all
policy-makers Lagrange multipliers $\mathbf{\mu}_{t}$. These algorithms then
add a step which is a change of vector basis for eliminating Lagrange
multipliers. Knowing the optimal path of variables ($\mathbf{y}_{t}$
$\mathbf{z}_{t}$), one can compute the Lagrange multipliers at the end of this algorithm:%

\begin{equation}
\mathbf{\mu}_{t}=\mathbf{P}_{y}\mathbf{y}_{t}+\mathbf{P}_{z}\mathbf{z}_{t}%
\end{equation}

\subsection{Step 5 (optional): An implementable representation of Ramsey
optimal policy}

Policymakers cannot implement a Ramsey optimal policy rule where policy
instruments responds to non-observable variables, such as the shocks
$\mathbf{u}_{t}$ or the Lagrange multipliers $\mathbf{\mu}_{t}$. They can
implement an observationally equivalent representation of the Ramsey optimal
policy rule where policy instruments responds to lagged observable variables,
including the lags of the policy instruments. This is also a useful
representation for testing Ramsey optimal policy using vector auto-regressive
system of equation.%

\begin{align*}
&  (H)\left\{
\begin{array}
[c]{c}%
\left(
\begin{array}
[c]{c}%
E_{t}\mathbf{y}_{t+1}\\
\mathbf{z}_{t+1}%
\end{array}
\right)  =\left(
\begin{array}
[c]{cc}%
\mathbf{A}_{yy}+\mathbf{B}_{y}\mathbf{F}_{y} & \mathbf{A}_{yz}+\mathbf{B}%
_{y}\mathbf{F}_{z}\\
\mathbf{0}_{zy} & \mathbf{A}_{zz}%
\end{array}
\right)  \left(
\begin{array}
[c]{c}%
\mathbf{y}_{t}\\
\mathbf{z}_{t}%
\end{array}
\right)  +\left(
\begin{array}
[c]{c}%
0\\
1
\end{array}
\right)  \mathbf{\varepsilon}_{t}\\
\mathbf{u}_{t}=\mathbf{F}_{y}\mathbf{y}_{t}+\mathbf{F}_{z}\mathbf{z}_{t}\\
\mathbf{x}_{0}=\mathbf{P}_{y,x}^{-1}\mathbf{P}_{y,kx}\mathbf{k}_{0}%
+\mathbf{P}_{y,x}^{-1}\mathbf{P}_{z,x}\mathbf{z}_{0}\text{, }\mathbf{k}%
_{0}\text{ and }\mathbf{z}_{0}\text{ given}%
\end{array}
\right. \\
&  \Leftrightarrow\left\{
\begin{array}
[c]{c}%
\left(
\begin{array}
[c]{c}%
E_{t}\mathbf{y}_{t+1}\\
\mathbf{u}_{t+1}%
\end{array}
\right)  =\mathbf{M}^{-1}\left(  \mathbf{A}+\mathbf{BF}\right)  \allowbreak
\mathbf{M}\left(
\begin{array}
[c]{c}%
\mathbf{y}_{t}\\
\mathbf{u}_{t}%
\end{array}
\right)  +\mathbf{M}^{-1}\left(
\begin{array}
[c]{c}%
0\\
1
\end{array}
\right)  \varepsilon_{t}\\
\mathbf{z}_{t}=\mathbf{F}_{z}^{-1}\mathbf{u}_{t}-\mathbf{F}_{z}^{-1}%
\mathbf{F}_{y}\mathbf{y}_{t}\\
\mathbf{x}_{0}=\mathbf{P}_{y,x}^{-1}\mathbf{P}_{y,kx}\mathbf{k}_{0}%
+\mathbf{P}_{y,x}^{-1}\mathbf{P}_{z,x}\mathbf{z}_{0}\text{, }\mathbf{k}%
_{0}\text{ and }\mathbf{z}_{0}\text{ given }%
\end{array}
\right.  \text{ }%
\end{align*}

where%

\begin{align*}
\mathbf{A}+\mathbf{BF}  &  \mathbf{=}\left(
\begin{array}
[c]{cc}%
\mathbf{A}_{yy}+\mathbf{B}_{y}\mathbf{F}_{y} & \mathbf{A}_{yz}+\mathbf{B}%
_{y}\mathbf{F}_{z}\\
\mathbf{0}_{zy} & \mathbf{A}_{zz}%
\end{array}
\right) \\
\left(
\begin{array}
[c]{c}%
\mathbf{y}_{t}\\
\mathbf{u}_{t}%
\end{array}
\right)   &  =\mathbf{M}^{-1}\left(
\begin{array}
[c]{c}%
\mathbf{y}_{t}\\
\mathbf{z}_{t}%
\end{array}
\right)  \text{ with }\mathbf{M}^{-1}=\text{ }\left(
\begin{array}
[c]{cc}%
\mathbf{1} & \mathbf{0}\\
\mathbf{F}_{y} & \mathbf{F}_{z}%
\end{array}
\right)
\end{align*}

In the estimation of dynamic stochastic general equilibrium model, the
controllable predetermined variables are usually set to zero at all periods.
They are as many auto-regressive forcing variables than controllable
forward-looking variables. If the number of policy instrument is equal to the
number of controllable forward-looking policy targets, $\mathbf{F}_{z}$ is a
square matrix which can be invertible. One eliminates forcing variables
$\mathbf{z}_{t}$ and replace them by policy instruments $\mathbf{u}_{t}$ in
the recursive equation, doing a change of vector basis. There is then of a
representation of forward-looking variables and policy instruments rule
optimal policy dynamics in a vector auto-regressive model. This representation
of Ramsey optimal policy rule is such that policy instruments $\mathbf{u}_{t}$
responds to lags of policy instruments $\mathbf{u}_{t-1}$ and of lags of the
observable policy targets $\mathbf{y}_{t-1}$. This representation can be
implemented by policy makers. It can be estimated by econometricians
(Chatelain and Ralf (2017a)). .

\subsection{Examples}

Chatelain and Ralf (2017a) use this algorithm for the new-Keynesian Phillips
curve as a monetary policy transmission mechanism. They check that it is
equivalent to Gali (2015) solution who used the method of undetermined
coefficients. They use the implementable representation of step 5 to estimate
structural parameters.

Chatelain and Ralf (2017b) use this algorithm for the new-Keynesian Phillips
curve and the consumption Euler equation as a monetary policy transmission
mechanism. They check the determinacy property of step 2 reduced form of the
Ramsey optimal policy rule.

Chatelain and Ralf (2016) use this algorithm for Taylor (1999) monetary policy
transmission mechanism. They check whether Taylor principle applies to Ramsey
optimal policy.

\section{Conclusion}

This algorithm complements Ljungqvist and Sargent (2012) algorithm taking into
account forcing variables. It is easy to code, check and implement.

\end{document}